\documentclass[11pt,english]{article}

\usepackage{graphicx}

\usepackage{amsfonts}
\usepackage{amsmath}
\usepackage[thmmarks]{ntheorem}
\newtheorem{theorem}{Theorem}[section]

\def\be{\begin{equation}}
\def\ee{\end{equation}}

\newtheorem{definition}{Definition}[section]
\newtheorem{propo}{Proposition}[section]
\theoremstyle{nonumberplain}
\theorembodyfont{\normalfont}
\theoremseparator{:}
\theoremsymbol{$\P$}
\newtheorem{demo}{Proof}

\begin{document}
\title{Simultaneity in Minkowski spacetime: from uniqueness to arbitrariness}
\author{Fabien Besnard,\\ EPF}

\maketitle
\abstract{In 1977, Malament proved a certain uniqueness theorem about standard synchrony, also known as Poincar\'e-Einstein simultaneity, which has generated many commentaries over the years, some of them contradictory. We think that the situation called for some cleaning up. After reviewing and discussing some of the literature involved, we prove two results which, hopefully, will help clarifying this debate by filling the gap between the uniquess of Malament's theorem, which allows the observer to use very few tools, and the complete arbitrariness of a time coordinate in full-fledged Relativity theory. In the spirit of Malament's theorem, and in opposition to most of its commentators, we emphasize explicit definability of simultaneity relations, and give only constructive proofs. We also explore what happens when we reduce to ``purely local'' data with respect to an observer.}

\section{Introduction}

How can an inertial observer in spacetime know that an event is simultaneous with  a particular point $O$ on its\footnote{To avoid any problem with gender, we have decided that the observer will be an asexual robot. Such an observer has the additional advantage of being possibly eternal, which is helpful in  mathematical arguments.} worldline ?

\begin{figure}[hbtp]
\begin{center}
\includegraphics[scale=0.8]{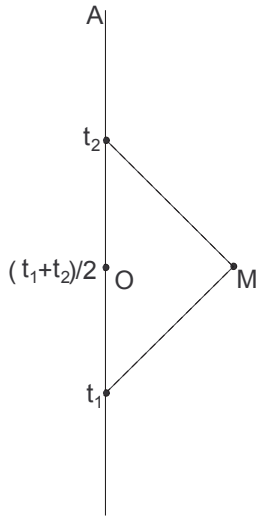}
\end{center}
\caption{Poincar\'e-Einstein simultaneity}
\end{figure}

Here is one way to proceed. At time $t_1$, as indicated by its wristwatch, the observer on the timelike straight line $X$ sends a light signal. The signal is reflected back to ${X}$ by a mirror at the spacetime point $M$, and then received by the observer at time $t_2$. The event $M$ is then assigned the time coordinate $t={1\over 2}(t_1+t_2)$. Let us call $O$ the event with time coordinate $t$ on $X$ (this event is physically defined by the position of the hand of the observer's watch). The procedure  defines the events $M$ and $O$ to be simultaneous.

What we have just described is the standard synchrony, also called the Poincar\'e-Einstein definition of simultaneity. For an inertial observer, this definition appears to be very natural, since it amounts to declaring simultaneous with $O$ all the events on the hyperplane which passes through $O$ and is orthogonal to the world-line $X$ with respect to  the Minkoswki metric. Since it has a geometrical meaning, one can wonder if it is really a convention. This is the problem of conventionality of simultaneity. Einstein himself clearly thought that simultaneity was a convention, as did Poincar\'e. This position has been fully developped and argued for in \cite{reich}, and became the more popular point of view for a while.  We will not review here the vast literature on this question, for which we refer to \cite{janis}. We will be only concerned with a sub-debate over a theorem put forward by Malament in \cite{malam}. This theorem has provoked a dramatic shift on the issue of conventionality, since contrarily to most expectations, it could be interpreted as saying that Poincar\'e-Einstein definition is \emph{not} a convention.

In section 2 we will review the papers of Malament \cite{malam}, Sarkar and Stachel \cite{sarkstach}, Giulini \cite{giul}, and Ben-Yami \cite{benyami}. We will also discuss the different kinds of structures involved.

In section 3 we will prove a result which can be viewed as an extension of Malament's theorem, adding the most natural piece of data to the hypotheses, namely a natural ordering on the world-line of the observer. This result  supplements some observations already made, in particular by Ben-Yami, by tackling with the issue of explicit definability, what had not been done since Malament's original theorem.

In section 4 we prove another extension, adding more hypotheses, which amount to allow the observer to use all the information provided by its clock. This result classifies some of the exotic equivalence relations quoted in \cite{giul}. Adding more structure would obviously allow the observer to define simultaneity using any coordinate function. Together, our two results can be seen as filling the gap between the non-conventionality of Malament's theorem, which allows the observer the observer to use very few tools, and the complete conventionality of full-fledged Relativity theory, which allows the observer to use any coordinates it likes.

To conclude we will see that things change radically if we remove $\kappa$ from the available data, and replace it with the ``local light cone structure'' of $X$, that is, the set of all light cones centered on a point in $X$. It will appear that, in this setting, the Poincar\'e-Einstein definition of simultaneity cannot be viewed as natural.

We will use the following notations and conventions. 

Throughout the text we use units in which $c=1$. The Minkowski spacetime will be denoted by ${\mathcal M}$. By this we mean that ${\mathcal M}=({\mathbb R}^4,\eta,Or)$, where $\eta$ is the Minkowski metric, and $Or$ is a time orientation. A time orientation can be defined thanks to a continuous timelike vector field, which allows to separate between future-directed and past-directed timelike vectors at any point (see \cite{beem}). Our signature convention is $(+,-,-,-)$. An event is a point in ${\mathcal M}$. For mathematical purposes, an inertial observer is just a timelike straight line. In the sequel, we fix such an inertial observer and call it $X$.

The relation of causal connectibility and lightlike connectibility between events $p$ and $q$ are respectively defined by :

\be
p\kappa q\Leftrightarrow \eta(p-q,p-q)\ge 0
\ee

and

\be
p\lambda  q\Leftrightarrow \eta(p-q,p-q)=0
\ee

The causal ordering $\preceq$ associated to the time orientation is defined by:

\be
p\preceq q\Leftrightarrow (p\kappa q\mbox{ and }q-p\mbox{ is future-directed})
\ee

The light cone (resp. forward light cone, resp. backward light cone, both of which can be defined thanks to the time orientation) of $p$ will be denoted $\Lambda_p$ (resp. $\Lambda_p^+$, resp. $\Lambda_p^-$). The ordering $\preceq$ and its dual $\succeq$ are completely characterized among partial order relations by the fact that the down sets $\downarrow p=\{x\in{\mathcal M}|x\preceq p\}$ are full half light cones. The choice of one these relations is equivalent to the choice of a time orientation.

Let $a\not= b$ any two points on $X$. The Poincar\'e-Einstein simultaneity relation with respect to $X$ is defined by

\be
p\Sigma_X q\Leftrightarrow \eta(p-q,a-b)=0
\ee

If $p\in X$, the hyperplane $\eta$-orthogonal to $X$ at $p$ will sometimes be denoted by $\Sigma_p$, with $X$ understood.

Let $p\in {\mathcal M}$. The projection of $p$ on $X$ is the unique point $\pi_X(p)$ on $X$ which satisfies $p\Sigma_X\pi_X(p)$ (that is, the event of $X$ which is simultaneous with $p$ according to the Poincar\'e-Einstein definition).

Let $v\in]0;+\infty[$  and $x\in X$. The set $\Gamma_v^+(x)$ is the set of events $y$ such that

\be
\eta(y-\pi_X(y),y-\pi_X(y))=-v^2\eta(\pi_X(y)-x,\pi_X(y)-x),\mbox{ and }x\preceq\pi_X(y)
\ee

The set $\Gamma_v^-(x)$ is similarly defined except that we require $x\succeq \pi_X(y)$. The sets $\Gamma_v^\pm(x)$ can be described as the family of world-histories of particles moving away from $x$  or towards it, depending on the sign,  at constant speed $v$, as measured by the inertial observer $X$. Note that these particles would be tachyons if $v>1$. The case $v=0$ is excluded by definition since it would not correspond to an equivalence relation. In the special case where $v=1$, we recover the backward and forward light cones of $x$. 

We can now define the \emph{general conic relation} of parameter $v$ with respect to $X$. This is the binary relation on ${\mathcal M}$ defined by :

\be
p\Gamma_{X,v}^\pm q\Leftrightarrow \exists x\in X,\ p\in \Gamma_v^\pm(x)\wedge q\in\Gamma_v^\pm(x)
\ee

In the particular case of $v=1$, we write $\Gamma_{X,1}^\pm=\Lambda_X^\pm$. In the limiting case $v=\infty$, we recover $\Sigma_X$.

\begin{figure}[hbtp]
\begin{center}
\includegraphics[scale=0.7]{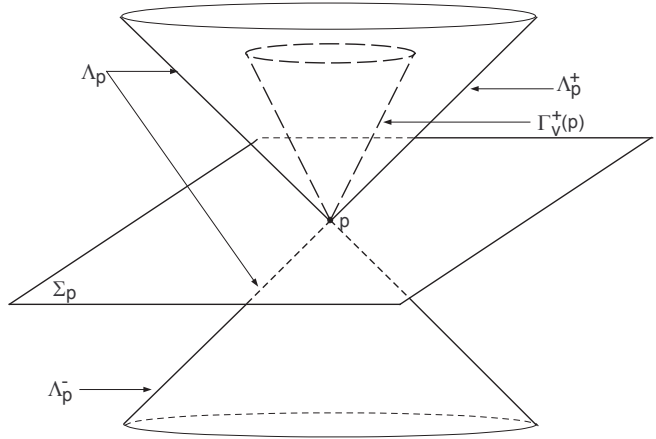}
\end{center}
\caption{Summary of the notations used throughout the text.}
\end{figure}

\section{Review and discussion}

We start of course with \cite{malam}, in which Malament proved the following theorem. Given an inertial observer $X$, the only binary relation on ${\mathcal M}$ which is:

\begin{enumerate}
\item an equivalence relation,
\item not trivial (all spacetime points are not equivalent, and at least a point on $X$ is equivalent to a point not on $X$),
\item invariant by all causal automorphisms stabilizing $X$,
\end{enumerate}

is the Poincar\'e-Einstein simultaneity relation relative to $X$. Let us recall that a causal automorphism is a bijection $f$ of Minkowski spacetime into itself, such that 

\be
p\kappa q \Leftrightarrow f(p)\kappa f(q)
\ee

This last requirement is motivated by the causal theory of time put forward by Gr\"unbaum. This is the setting in which Malement's discussion takes place (note that Malament does not endorse the named theory). In such a viewpoint, a simultaneity relation is not conventional if it can be defined by a first-order fomula using only the causal connectibility relation and the given world-line $X$. Malament first proved that Poincar\'e-Einstein simultaneity can be so defined\footnote{For completeness, we give the proof again in section 3.}, and then established the unicity result stated above. Since a relation which is definable by a first-order formula using the mentioned data is evidently invariant under causal automorphisms stabilizing $X$,  Poincar\'e-Einstein simultaneity is the only such relation.

Here Malament introduced a vocabulary that is perhaps misleading by distinguishing \emph{explicitly} definable (that is, by a first-order formula), and \emph{implictly} definable (which really means invariant under automorphisms). This aroused some confusion in the debate to come. Indeed, most commentators of Malament focused on implicit definability, and proved results which are unnecessarily weak. We will settle this issue in this paper, but for the moment, let us write down a (not very formal) definition, for future reference.

\begin{definition} Let $S$ be some structure on the set ${\mathbb R}^4$, and $R$ be a relation on the same set. We say that:
\begin{enumerate}
\item $R$ is explicitly $S$-definable if, and only if, $xRy\Leftrightarrow \Phi(x,y)$ where $\Phi(x,y)$ is a first-order formula with free variables $x,y$, perhaps some bound variables, and references to $S$.
\item A bijection $f : {\mathbb R}^4\rightarrow {\mathbb R}^4$ which is an automorphism for the structure $S$ will be called an $S$-automorphism. We write ${\rm Aut}_S$ for the set of all $S$-automorphisms.
\item $R$ is implicitly $S$-definable if, and only if, $R$ is ${\rm Aut}_S$-invariant. That is: $\forall f\in{\rm Aut}_S$, $\forall x,y\in{\mathbb R}^4$, $xRy\Rightarrow f(x)Rf(y)$.
\end{enumerate}  
\end{definition}

Some remarks are called for.

\begin{itemize}
\item Experts in logic might find the definition of explicit definability too vague. However, it will suffice for what we intend to do in this paper.
\item More precise definitions (several kinds of them in fact) can be found in \cite{ryn}. In any case, it is clear that explicit definability entails implicit definability, and this is in fact all we really care about in this text.
\item What is an $S$-automorphism will be made explicit in all the cases considered below. The definition above is only intended for notational purposes.
\item In this definition, we consider ${\mathbb R}^4$ instead of ${\mathcal M}$, because ${\mathcal M}$ already comes with a structure (namely an affine structure, a metric, and a time orientation), that we  do not always want to use. \emph{It is only the ``naked'' set} ${\mathbb R}^4$ that we always take for granted.
\item It is known (see for instance \cite{malam}) that the affine structure of ${\mathcal M}$ and the light cones are $\kappa$-definable. We will freely use these facts below and recall the proofs in section 3.
\item The implication symbol in $xRy\Rightarrow f(x)Rf(y)$ can be replaced by an equivalence, since $f^{-1}$ is also an $S$-automorphism.
\end{itemize}

The biggest source of confusion in the debate over Malament's theorem is a paper of Sarkar and Stachel, \cite{sarkstach} in which these authors criticized the theorem on the ground that the half light cones are definable from the causality relation alone. Here goes the argument: let $e$ be a point in spacetime and $\Lambda_e$ the light cone of $e$. Two points on $\Lambda_e$ are on the same half cone emanating from $e$ iff they are not causally connected, or they are connected by a signal that neither contains $e$ nor any point in the elsewhere of $e$. We can note that there is a simpler caracterization: using the convexity of a full half-cone, and the non-convexity of the full light cone, we have:

\be
p\mbox{ and }q\mbox{ belong to the same half-cone}\Leftrightarrow \forall m\in]pq[, m\kappa e\mbox{ and }m\not=e
\ee

This permits to define the two half light cones from $e$. Then, Sarkar and Stachel say, we can arbitrarily choose one to be $\Lambda_e^+$ and the other to be $\Lambda_e^-$ and parallel transport this choice to any other spacetime point. The appeal to parallel transport is suspicious, but in fact it is not necessary to use this tool, for we can assure that all the choices are coherent along the given world-line $X$, which is all we care about, by requiring that for any events $e$ and $e'$ on $X$, one has $\Lambda_e^+\cap\Lambda_{e'}^-$ or $\Lambda_e^-\cap\Lambda_{e'}^+$ non empty. Of course half-cones are excluded by Malament's postulate because they are not invariant under time reflections. Thus, Sarkar and Stachel argued that the causal automorphisms used in the proof should be required to preserve time orientation. A ``natural'' way of achieving this is to stipulate that the causal automorphisms should be connected to the identity. Then, they automatically preserve space as well as time orientation.  They proposed to prove the following:

\begin{theorem}\label{sstheorem}
\begin{enumerate}
\item Poincar\'e-Einstein simultaneity is the only non-trivial (in the sense of Malament) ${\rm Aut}_{X,\preceq}$-invariant relation that depends only on the inertial frame to which $X$ belongs, and not on the particular worldline $X$.
\item The relations defined by the past (resp. future) light cones are the only non-trivial ones which are ${\rm Aut}_{X,\preceq}$-invariant and satisfy the following additional condition: given  another inertial observer $X'$ such that $X\cap X'=\{e\}$, the equivalence class of $e$ is the same for $\Lambda_X^\pm$ and $\Lambda_{X'}^\pm$.
\end{enumerate} 
\end{theorem}

This theorem needs some explanations. First, we recall that an inertial frame is a foliation of spacetime by parallel straight lines. Note also that we have rephrased their initial claims in accordance with the notations used in this paper. Finally, we have stated the theorem with ${\rm Aut}_{X,\preceq}$, which is the group of automorphisms preserving $\preceq$ and $X$, instead of the group of causal automorphisms preserving spatial and temporal orientation. We felt free to make this change because it only amounts to removing the space reflections, which are of no use whatsoever in the proof. In this way we avoid unnecessary distinctions.

Unfortunately, there are several problems with their arguments. First,  we cannot agree with them when they criticize Malament's result. The relations given by the past (resp. future) half-cones do not depend only on the causal structure, but also on the choice of a time orientation. This point has been raised and discussed at length in \cite{ryn}. The source of the problem is of course the statement ``we can arbitrarily choose one to be $\Lambda_e^+$'', because this choice is precisely equivalent to the prescription of a time orientation in Minkowski space.  

Finally, another problem, this one in the proof of theorem \ref{sstheorem}, has been identified by Giulini in \cite{giul}, and we now turn our attention to this paper. First, Giulini criticized Sarkar and Stachel proof for its dependence on the use of scale transformations which are no more symmetries of physics than time reflections. He thus proposed to use the proper orthochronous Lorentz group as automorphism group. Secondly, he noticed a shortcoming in the proof of the first claim of theorem \ref{sstheorem}. While Giulini's remark is right, this theorem is true nonetheless, and will be recovered as an immediate corollary of theorem \ref{montheorem1} in section 3.

Giulini proposed to clarify the situation by proving the following:

\begin{theorem}\label{giultheorem}
Let $\Xi$ be an inertial frame and $S$ a non-trivial\footnote{Here non-trivial means different from the identity and the universal relation.} ${\rm Ilor}_\Xi$-invariant equivalence relation on ${\mathcal M}$, where ${\rm Ilor}$ is the proper orthochronous Lorentz group. Then the possible equivalence classes $[p]$ are given by:
\begin{enumerate}
\item The hyperplane $\Sigma_p$ which is $\eta$-orthogonal to $\Xi$ and contains $p$. 
\item The union over $n\in{\mathbb Z}$ of $\tau^n(\Sigma_p)$ where $\tau$ is a translation parallel to the lines of $\Xi$.
\item The line $X\in\Xi$ which contains $p$.
\item The union  over $n\in{\mathbb Z}$ of $\tau^n(p)$ where $\tau$ is as above.
\end{enumerate}
\end{theorem}

Since Malament's initial soundness criterion is here insufficient to discard some obviously unwanted cases, Giulini introduced another, more stringent, criterion: each equivalence class must intersect any physically realizable trajectory in at most one point. Looking at the list above, it is easy to see that:

\begin{theorem}\label{giultheorem2} Poincar\'e-Einstein simultaneity is the unique ${\rm Ilor}_\Xi$-invariant equivalence relation which satisfies Giulini's soundness condition.
\end{theorem}

Giulini also remarks that if one replaces ${\rm Aut}_{X,\kappa}$ by ${\rm Ilor}_X$, the group of proper orthochronous Poincar\'e transformations stabilizing $X$, in the hypotheses of Malament's theorem (this amounts to removing scale transformations from the automorphism group), then the Poincar\'e-Einstein relation is no longer unique. He also quotes some examples which are sufficiently pathological to demonstrate that there is no hope to generalize Malament's theorem in a sensible way with these new hypotheses. However, we will show in this paper (section 4) that if we tighten the soundness condition in a very natural way, we can not only give a classification of these equivalence relations, but also prove that they are all explicitly definable in terms of the given structure.

It should be noted that, while Giulini has enriched the debate with more rigourous statements and proofs than those of Sarkar and Stachel, they depart somewhat from the setting in which Malament deliberately placed himself, that is Gr\"unbaum's causal theory of time. We do not think that criticizing Malament's requirement of invariance under scale transformations and time reflection on the ground that their are not physically realizable is  completely to the point. Invariance under these transformations is a \emph{necessary condition} given that the simultaneity relation is supposed to be definable from $\kappa$ and $X$. This hypothesis, in turn, is presented by Malament as  natural in the setting of the causal theory of time. Thus,  the critics of Malament's theorem should rather challenge the idea that a simultaneity relation has to be definable thanks to $\kappa$ alone than arguing against its logical consequences. Such a criticism has been carried out by Ben-Yami in \cite{benyami}. This author has argued that even in the context causal theories of time, one should use simultaneity relations defined thanks to the causal ordering\footnote{Ben-Yami uses the notation $\sigma$ instead of $\preceq$.} $\preceq$, instead of the (symmetric) causal connectibility relation $\kappa$. Ben-Yami then proves that the simultaneity relations implicitly definable from $\preceq$ and $X$, that is, the ${\rm Aut}_{X,\preceq}$-invariant equivalence relations, which satisfy Malament's soundness condition\footnote{In fact, he uses both Malament and Giulini criteria, but Malement's one suffices.}, are all the `conic relations'' of the type $\Gamma_{X,v}^\pm$  including $\Lambda_X^\pm$ as particular cases, and $\Sigma_X$ as a limiting case.  Ben-Yami cites Spirtes \cite{spirtes} as the first author to have noticed that an infinity of simultaneity relations appears if one discards time reflections from the automorphism group. Then Ben-Yami introduces a soundness condition of his own, which is stated in the following way ``If event $e_1$ is a cause of event $e_2$, then $e_2$ does not precedes $e_1$''. Technically this can be formulated like this: let $<$ stands for the relation defined by $e<e'\Leftrightarrow \exists x,x'\in X$, $x\not=x'$ such that $x$ is simultaneous with $e$, $x'$ with $e'$, and $x\preceq x'$. Then $e_2$ precedes $e_1$ means $e_2<e_1$. Thus Ben-Yami's soundness condition is:

\begin{equation*}
e_1\preceq e_2\Rightarrow \neg(e_2<e_1)\tag{BY}
\end{equation*}
 
With this condition there only remain the conic relations with cones outside the light cone (including the limiting cases of $\Lambda_X^\pm$ and $\Sigma_X$). Ben-Yami also worries about explicit definability, though he wrongly credits Rynasiewicz for this notion, whereas it was already in \cite{malam}. He notes that $\Lambda_X^\pm$ are explicitly definable from $X$ and $\preceq$ and gives a formula. However, he is not exhaustive here, since, as we will see, every other conic relation is also definable in this sense.

At this point a summary of what has been proved will probably be helpful. It is given in the following table.

\begin{table}[hbtp]
\begin{minipage}{12cm}
\begin{tabular}{|c|*{4}{p{2.3cm}|}c|}
\hline
Structure&Classifies  explicitly definable eq. relations&Classifies implicitly definable eq. relations& Soundness condition &Result of classification & Proved in\\
\hline
$(\eta,Or)$ & No & Yes & None & $I$, $U$ & \cite{giul}\\
\hline
$(X,\kappa)$ & Yes & Yes & M & $U$, $\Sigma_X$ & \cite{malam}\footnote{Alternative proof of the implicit part in \cite{giul}.}\\
\hline
$(\Xi,\eta,Or)$ & No & Yes & G & $\Sigma_X$ & \cite{giul}\\
\hline
$(X,\preceq)$ & No & Yes & M,G & $\Gamma_{X,v}^\pm$, $v\in]0;\infty]$ & \cite{benyami}\\
\hline
$(X,\preceq)$ & Yes & Yes & M & $\Gamma_{X,v}^\pm$, $v\in]0;\infty]$ & This paper\\
\hline
$(X,\eta,Or)$ & Yes & Yes & B & $\Gamma_{X,\tilde f}$ & This paper\\
\hline
\end{tabular}
\end{minipage}
\end{table}

In the table above, $I$  stands for the identical relation $xIy\Leftrightarrow x=y$, $U$ for the universal relation ($xUy$ always true), and a $Or$ is a time orientation. The relation $\Gamma_{X,\tilde f}$ will be defined in section \ref{secth2}.

Note that in \cite{giul}, the relations which are classified are required to be invariant under ${\rm Ilor}$ or a subgroup of ${\rm Ilor}$, so that we should logically have put a space orientation in the given structure. However, the space orientation is not used in the proof (and the result shows indeed that the relations that are found are also invariant under space reflections).

Remember the following soundness conditions :

\begin{equation*}
R\not=U\mbox{ and }\exists x\in X, \exists y\notin X, xRy\tag{M}\label{malamcond}
\end{equation*}

\begin{equation*}
R\not=I,\mbox{ and no two points on a physically realizable trajectory are equivalent}\tag{G}
\end{equation*}

We add the following one:

\begin{equation*}
\forall y\in{\mathcal M},\exists!x\in X, xRy\tag{B}\label{mycriterion}
\end{equation*}
 
A word must be said on the motivations behind the choice of the various structures allowed. We have already noted that Malament's initial choice of $\kappa$ was dictated by Gr\"unbaum's causal theory of time, and we will not debate this issue here. However, from a physical point of view, if one seeks to break Malament's conclusion, adding a time orientation to the data seems the most natural step. Indeed, it is well known that Nature does distinguish between the four different components of the Poincar\'e group. However, we do not wish to import knowledge from particle physics, invoking time asymetric particle productions as Sarkar and Stachel, do in order to motivate the introduction of a time orientation. Instead, we take an operational point of view. In the process described at the  beginning of this paper, the observer uses a clock. A clock provides $X$ with three different things, which are three different aspects of time that Relativity taught us to clearly distinguish: a chronology, an ordering, and a measure of durations. All these are only defined on $X$ to begin with. We will represent a chronology by a bijective function $f : X\rightarrow {\mathbb R}$, an ordering by a total order relation $<$ on $X$, and a measure of durations by a Borel measure $\mu$ on $X$ such that $\mu([xx'])=\sqrt{\eta(x-x',x-x')}$.  Arbitrary $f$ or $<$ will  not qualify as proper chronology or ordering. We require that $f$ imports the topology of the real line and thus be an homeomorphism. As for $<$, we demand it to satisfy the naturalness criterion below: 

\begin{definition}
An order relation $<$ is \emph{natural} if for all $x\in X$, the down-set $\downarrow x=\{x'\in X|x'<x\}$ (that is, the past of $x$), is an open half-line.
\end{definition}

Take two homeomorphisms $f_1,f_2 : X\rightarrow {\mathbb R}$. Then $f_1\circ f_2^{-1}$ is a continuous bijection from ${\mathbb R}$ to itself, and as such must be monotone. Define $f_1\sim f_2$ if $f_1\circ f_2^{-1}$ is increasing. Then there are two classes for this equivalence relation on homeomorphisms from $X$ to ${\mathbb R}$. Similarly, there are exactly two natural orderings of $X$. The choice of such a class (denoted $[f]$ in the sequel) or ordering, is equivalent to an orientation of $X$.

A clock carried by $X$ may be modelized by continuous function $T : X\rightarrow {\mathbb R}$ such that  $(T(x)-T(x'))^2=\eta(x-x',x-x')$.  It is clear that two clocks differ only by a sign and an additive constant. If we choose to ignore the choice of origin, then we work with a class of clocks up to an additive constant, and we write $\tilde{T}$. Since the hypotheses imply that $T$ is an homeomorphism, it defines a chronology. It also singles out a unique Borel measure $\mu$ on $X$ such that $\mu([xx'])=|T(x)-T(x')|$. The theorem below shows, among other things, that $[T]$ defines a natural ordering on $X$.

\begin{theorem}
Let us note $A\leftrightarrow B$ if  $A$ can be defined thanks to $B$ and $B$ to $A$. Then, using the same notations as above, we have
$$ (X,[f]) \leftrightarrow  (X,<)$$

$$(X,\kappa,<) \leftrightarrow (X,\preceq)\leftrightarrow (X,\kappa,Or)$$

and 
$$(X,\kappa,\tilde{T})\leftrightarrow (X,\eta,Or)$$
\end{theorem}
\begin{demo}
The proof is easy and we only sketch it.

First equivalence : given $[f]$,  we define $<$ by $x<x'\Leftrightarrow f(x)<f(x')$. If $g\sim f$, then it is easy to see that $g$ defines the same ordering. Let $x\in X$. Then the down-set $\downarrow x$ is $\{x'\in X|f(x')<f(x)\}=f^{-1}(]-\infty;f(x)[$. Since $f$ is a homeomorphism, $\downarrow x$ is an open half-line (open, connected, unbounded). Conversely, given $<$, an homeomorphism from $X$ to ${\mathbb R}$ must send open half-lines to open half-lines, and thus must be increasing or decreasing. The increasing homeomorphisms from $X$ to ${\mathbb R}$ form an equivalence class of $\sim$.

Second equivalence : Suppose we are given $(X,\kappa,<)$. A causal ordering on ${\mathcal M}$ is a (partial) order relation $\preceq$ such that the down-sets $\downarrow x$ are (full) half light cones. Transitivity shows that there are only two such relations. Let $x,y\in{\mathcal M}$, and call $x_1,x_2$ (resp. $y_1,y_2$) the intersection points of $X$ with $\Lambda_x$ (resp. $\Lambda_y$) such that $x_1<x_2$ (resp. $y_1<y_2)$. We take $x_1=x_2$ (resp. $y_1=y_2$) in case $x$ (resp. $y$) is on $X$. Then the relation $x\preceq y\Leftrightarrow (x_1\leq y_1$ and $x_2\leq y_2)$ is easily seen to be a causal order. Now if $\preceq$ is a causal order, $\kappa$ is easily recovered. Take $x$ and $x'$ on $X$ such that $x\preceq x'$ and $x\not=x'$. Define the constant vector field $\xi$ on ${\mathcal M}$ by $\xi(p)=x'-x$. Then $\xi$ is a continuous timelike vector field, which defines a time orientation.  Finally, from a continuous timelike vector field $\xi$, we define a natural order on $X$ by $x<y\Leftrightarrow y$ belongs to the open half-line $]x;x+\xi(x))$ (open half-lines are $\kappa$-definable, see section 3).

Third equivalence : It is well known that $\kappa$ determines the conformal class of $\eta$, so that $T$ fixes the remaining degree of freedom and permits to recover $\eta$ from $\kappa$ and $T$. However, we prefer to give the explicit construction, which is surprisingly simple. Fix an origin $o$ on $X$. Using the polarization identity, it suffices to define $\eta(p-o,p-o)$ for any $p\in{\mathcal M}$ in order to completely determine $\eta$. Then, if $x_1,x_2$ are the two intersection points of $\Lambda_p$ and $X$ (possibly with $x_1=x_2=p$), then $\eta(p-o,p-o)=T(x_1)T(x_2)$. Moreover, since we evidently have $(X,\kappa,T)\rightarrow (X,\kappa,<)$, we can define $Or$ using the second equivalence. Conversely, we obviously have $\eta\rightarrow \kappa$. Using the second equivalence, we see that $(X,\eta,Or)\rightarrow (X,\kappa,Or)\rightarrow (X,<)$. Using this $<$, and choosing an origin $x_0$, we can define a function $T$ by $T(x)=\sqrt{\eta(x-x_0,x-x_0)}$, if $x_0\leq x$, and $T(x)=-\sqrt{\eta(x-x_0,x-x_0)}$ is $x<x_0$. Then $\tilde{T}$ does not depend on $x_0$.
\end{demo}

The first of these equivalences tells us that having a (natural, as defined above) notion of past at each point of $X$ is equivalent to having a ``bad clock'', that is a clock which can run unevenly but is required to never go backwards.  If we only care about the direction of time, any time asymetric phenomenon, including the formation of memories in a human brain, can serve as a ``bad clock''. This can be important in some philosophical debates about the existence of the present (see \cite{bes}). 

Of course, Malament's theorem shows that we do not really need a clock to define $\Sigma_X$, even a bad one. Indeed, we only need to define the midpoint of a segment $[xx']$ on $X$, and this can be done using $\kappa$ alone. The midpoint $m$ of $[xx']$ is the only point of $X$ for which the following formula is true:

\be
\exists a,b\in\Lambda_x\cap\Lambda_{x'}, \Lambda_a\cap\Lambda_b\cap\Lambda_m=\emptyset
\ee\label{tru}

However, we can ask to what extent this definition is operational. How can $X$ know that the three lightcones $\Lambda_a,\Lambda_b,\Lambda_m$ have no common point ? More generally, how can $X$ be informed that $p\kappa q$ if neither $p$ nor $q$ lie on $X$ ? Well, he or she cannot. We will have more to say about this later. For the moment, we will accept $\kappa$ as ``God-given'', and proceed to show the two last theorems of the table, which correspond to the progressive enrichment of the given structure, from $(X,\kappa)$ (Malament), to $(X,\kappa,<)$, and then $(X,\kappa,\tilde{T})$.

\section{Theorem 1}

In the construction below we will need to use half-lines, segments, and light cones. All of these objects are explicitly $\kappa$-definable.

\begin{propo}\label{propo1} If no two points among $a,b,c$ are causally connectible, then $a,b,c$ are colinear if, and only if, $\Lambda_a\cap\Lambda_b\cap\Lambda_c=\emptyset$.
\end{propo}
\begin{demo} (sketch) 
First, suppose $a,b,c$ are colinear and $\exists m\in{\mathcal M}$, $a,b,c\in\Lambda_m$. Then $a,b,c$ are all on a light ray passing through $m$, and so are causally connectible. This shows the direct part.

For the converse, suppose $\Lambda_a\cap\Lambda_b\cap\Lambda_c=\emptyset$ and $a,b,c$ are not colinear. Hence they define a $2$-plane. Suppose this $2$-plane is spacelike. Then there exists a point $o$ which is equidistant (for the the Riemannian metric induced by $\eta$) from $a,b,c$. Intuitively speaking, light take the same time to go from $o$ to $a,b,c$, so that the three light cones $\Lambda_a$, $\Lambda_b$, $\Lambda_c$ meet. We let reader write a formal proof. If the $2$-plane $(abc)$ is Lorentzian, it is not so obvious that the three light cones meet, but can be shown by direct calculation using well chosen coordinates.
\end{demo}

\begin{propo}\label{propocol}
If $a\kappa b$, and $\neg(a\lambda b)$, then $a,b,c$ are colinear if, and only if, there is no point $m$ such that $m$ belongs to three lines $(aa')$, $(bb')$, $(cc')$ with $a,a'\in\Lambda_b\cap\Lambda_c$, and $a\not=a'$, and similarly for $b,b',c,c'$.
\end{propo}
\begin{demo}
First note that the $m\in(aa')$ is $\kappa$-definable thanks to proposition \ref{propo1}, and similarly for the other lines appearing above. It is easy to see that if $a,b,c$ are colinear and causally connectible, but not lightlike with respect to each other, the respective intersection of two of the three light cones $\Lambda_a$, $\Lambda_b$, $\Lambda_c$ are in parallel hyperplanes, thus no $m$ as the statement of the proposition exists. Conversely, if $a,b,c$ are not colinear, they define a Lorentzian $2$-plane. In this case, the point $m$ is easily seen to exist by elementary geometry, as shown in figure \ref{figpropo}.

\begin{figure}[hbtp]
\begin{center}
\includegraphics{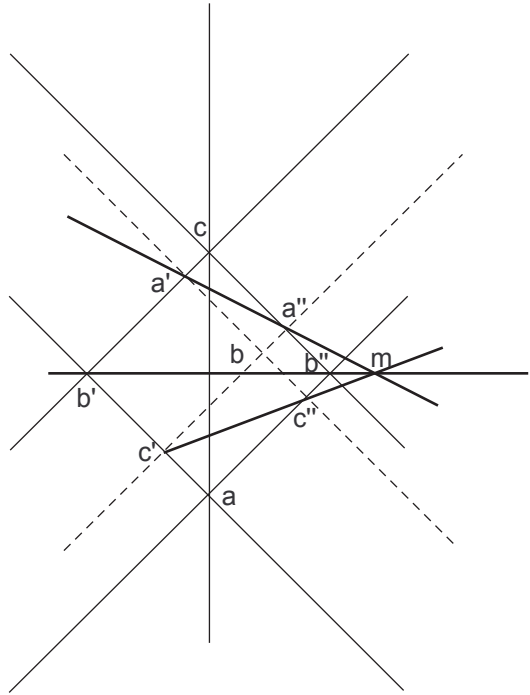}
\caption{Figure for proposition \ref{propocol}}\label{figpropo}
\end{center}
\end{figure}
\end{demo}

There remains the case where two points are in the same light cone.

\begin{propo}
If $p\lambda q$ and $p\not=q$, then $r\in[pq]$ if, and only if, $(x\kappa q$ and $x\kappa p)\Rightarrow x\kappa r$.
\end{propo}
\begin{demo}
We can suppose without loss of generality that $p\preceq q$. 

Let us first prove the direct part. Let $r\in[pq]$ and $x$ be such that $x\kappa q$ and $x\kappa p$. Thus $p\preceq r\preceq q$. First case: $q\preceq x$. Then by transitivity of $\preceq$, $r\preceq x$, thus $r\kappa x$. Second case: $p\preceq x$. This is similar to the first case. Third case: $p\preceq x\preceq q$. Then $x\in[pq]$, and clearly $x\kappa r$.

Now for the converse. If $x\in[pq]$, then $x\kappa q$ and $x\kappa p$, hence we must have $x\kappa r$. This shows that the full light cone of $r$ contains $[pq]$, which is lightlike. Thus $r\in(pq)$. It is then easy to see that we must have $r\in[pq]$.
\end{demo}

Of course, this proposition allows to define the half-line $[pq)$ as the set of points $r$ such that $r\in[pq]$ or $q\in[pr]$, and the line $(pq)$ as $[pq)\cup[qp)$. Though we will not need this fact, it is easy to use lightlike segments to define timelike and spacelike ones by projecting with light cones as in the figure below.

There remains to show that the light cone $\Lambda_e$ is $\kappa$-definable. This will also show that $\lambda$ is $\kappa$-definable\footnote{The converse is true but is not needed in the sequel.}. For this we just need a converse of the proposition above. Hence, the formula (which is given in \cite{malam}) is the following:

\begin{propo}
If $p\not=q$, $p\lambda q\Leftrightarrow (p\kappa q\wedge(\forall r\in[pq],\ \forall x\in{\mathcal M},\ x\kappa p\wedge x\kappa q\Rightarrow x\kappa r))$.
\end{propo}
 
We just need to show the $\Leftarrow$ part. Instead of a formal proof, we provide a figure, which is clear enough.

\begin{figure}[hbtp]
\includegraphics{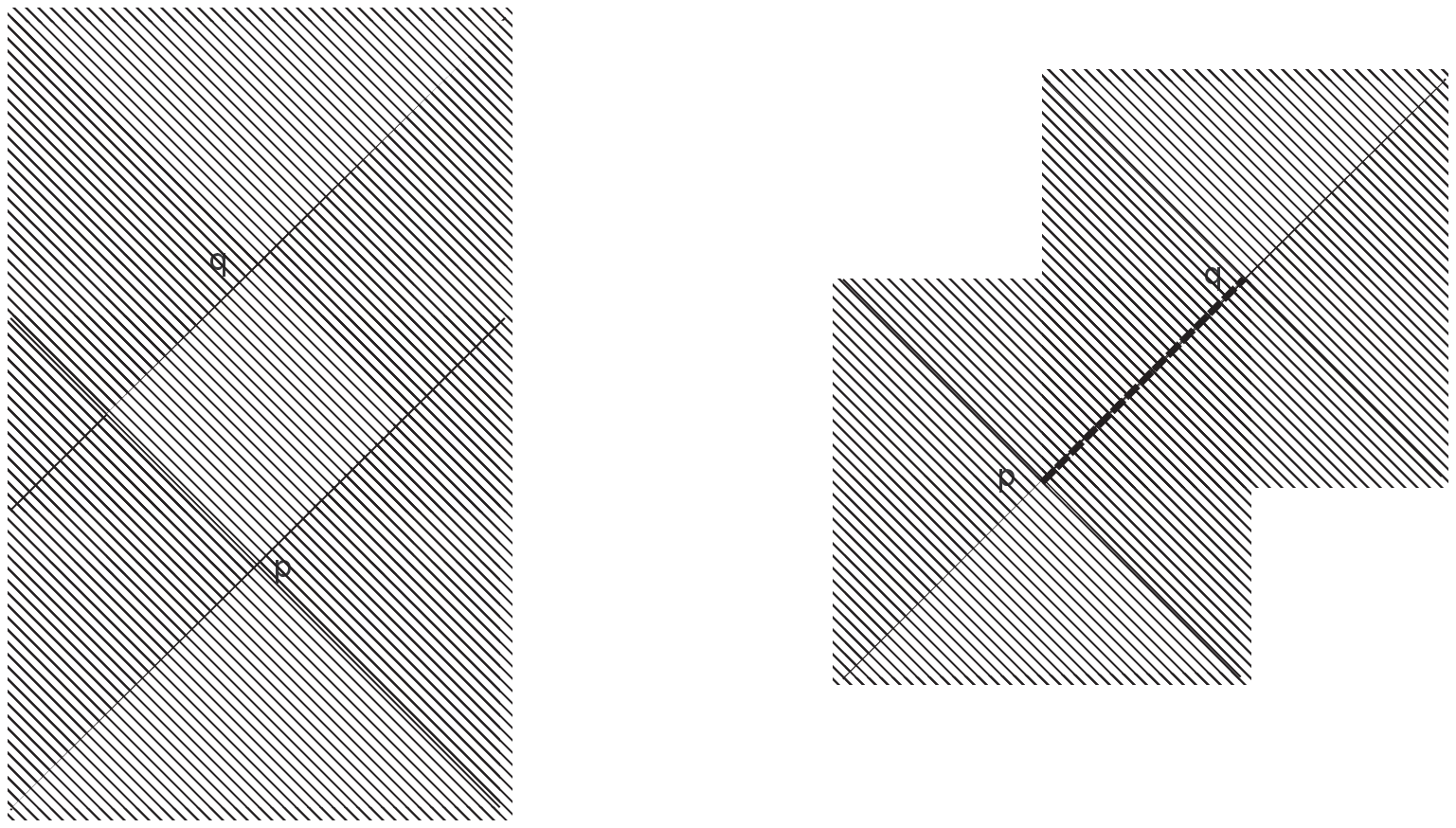}
\caption{The intersection of the full light cones of $p$ and $q$, in case $p\kappa q$ and $\neg(p\lambda q)$ (to the left), and in case $p\lambda q$ (to the right).}
\end{figure}

\begin{theorem}\label{montheorem1}
Let $X$ be an inertial observer and $<$ a natural ordering on $X$. If $R$ is a relation on Minkowski spacetime such that $R$ 
\begin{enumerate}
\item is an equivalence relation,
\item satisfies (\ref{malamcond}),
\item is implicitly $(X,\kappa,<)$-definable,
\end{enumerate}
then  $R$ is either   $\Sigma_X$, the Poincar\'e-Einstein simultaneity relation with respect to $X$, or a relation of the form $\Gamma_{X,v}$, with $v\in{\mathbb R}\setminus\{0\}$. Conversely, all these relations are \emph{explicitly} $(X,\kappa,<)$-definable.
\end{theorem}
\begin{demo}
First, let $R$ be a relation satisfying all the hypotheses. 

Let $f$ be either:
\begin{itemize}
\item a proper orthochronous Poincar\'e transformation which fixes all elements of $X$ (that is, a transformation of the form ${\rm Id}_X\oplus r$, where $r$ is a (linear) rotation of ${\mathbb R}^3$),
\item a translation in the direction of $X$,
\item a scale transformation about a point $o\in X$, that is a transformation of the form $(t;\vec{x})\mapsto \alpha(t-t_o;\vec{x}-\vec{x}_o)+(t_o;\vec{x_o})$, where $\alpha>0$ and $O=(t_o;\vec{x_o})$ in appropriate inertial coordinates,
\end{itemize}

then $f$ is a one-to-one map such that $x\kappa y\Leftrightarrow f(x)\kappa f(y)$, $f(X)=X$, and for every points $x,y$ on $X$, $x<y\Leftrightarrow  f(x)<f(y)$. Thus, $f\in{\rm Aut}_{X,\kappa,<}$. Note that, conversely, every element of ${\rm Aut}_{X,\kappa,<}$ is a combination of these ones, but we will not need this fact. 

Now since $R$ satisfies (\ref{malamcond}), let $x\in X$ and $y\notin X$ be such that $xRy$. Let us write $[x]$ for the class of $x$ with respect to $R$.

\begin{enumerate}
\item First case: $y$ belongs to the hyperplane $\Sigma_x$ $\eta$-orthogonal to $X$ at $x$ (that is, $x\Sigma_X y$).

By first using scale transformations about $x$, we see that the whole half line $[xy)$ belongs to $[x]$. Then, by using all the automorphisms of the form ${\rm Id}_X\oplus r$, with $r$ a rotation in $\Sigma_x$ about an axis containing $x$, we see that $\Sigma_x\subset [x]$.

If we now use translations in the direction of $X$, we see   that for all $e\in X$, $[e]$ contains the hyperplane $\Sigma_e$ orthogonal to $X$ at $e$. Suppose that $e'\not=e$, $e'\in X$, and $\Sigma_{e'}\subset [e]$. Then by using a scale transformation about $e$, we have $\Sigma_z\subset [e]$ for all $z\in X$, thus $R$ is trivial. We conclude that the hyperplanes $\Sigma_e$ are all in different classes, and this shows that $R=\Sigma_X$.
\item Second case: $y$ is not in $\Sigma_x$. Then, using scale transformations and rotations again, we see that the half cone generated by rotating the half line $[xy)$ around $X$, that is $\Gamma_v^\pm(x)$, belongs to $[x]$.  Using the same technique as above (translation in the direction of $X$ and scale transformations), we easily show that each equivalence class contains a unique such half cone.

\end{enumerate}

Now for the converse. It is known (see for instance \cite{malam}) that $\Sigma_X$ is explicitly $(X,\kappa)$-definable. We can give another proof using formula (\ref{tru}): $p\Sigma_Xq$ if and only if $[p_1p_2]$ and $[q_1q_2]$ have the same midpoint, where $\{q_1;q_2\}=\Lambda_q\cap X$, and similarly for $\{p_1;p_2\}$.

It is also easy to see that the relations defined by the past light cones or the future light cones, respectively, are explicitly $(X,\kappa,w)$-definable.  For instance, one has:

$$x\Lambda_X^+y\Leftrightarrow (\exists e\in X, x,y,\in\Lambda_e)\wedge(\exists x'\in X\cap\Lambda_x, e<x')\wedge(\exists y'\in X\cap \Lambda_y, e<y')$$

Since the light cones are definable in terms of $\kappa$, this proves the claim. The case of $\Lambda_X^-$ is of course similar.

However, it is perhaps less obvious that any ``conic'' relation is definable in terms of $\kappa$, $X$, and $<$. To see that this is possible, let us first show how to build a square lattice in a particular $2$-plane, using tools which obviously depend only on $\kappa$, $X$ and $<$.

First, let us consider $x_{00}\in X$ and $x_{10}\in\Lambda_{00}^+$ (where we write $\Lambda_{00}^+$ instead of $\Lambda_{x_{00}}^+$). We will have to remember later that the  construction below depends on the choice of these two points.

We define $x_{01}$ to be the intersection point of $\Lambda_{10}^+$ and $X$. Thus $x_{00}$, $x_{10}$ and $x_{01}$ are defined.

Let $n\geq 1$ and suppose $x_{n,0},x_{n-1,0},\ldots,x_{0n}$ are already defined. Then we define $x_{n+1,0}$ to be the only point on the line  $(x_{00}x_{10})$ which satisfies $x_{n+1,0}\Sigma_X x_{n-1,1}$ (see figure \ref{quadrillage}). We now define $x_{0,n+1}$ to be $\Lambda_{n+1,0}^+\cap X$. Then for all $k$ such that $1\leq k\leq n$, we let $x_{n+1-k,k}=\Lambda_{n-k,k}^+\cap (x_{n+1,0}x_{0,n+1})$.

We let the reader convince himself/herself that it is easy to devise a similar procedure to fill the whole half-plane with the lattice, starting with $x_{2,-1}\in (x_{10}x_{01})$ and such that $x_{2,-1}\Sigma_X x_{00}$, using backward half light cones, and lines already constructed.

\begin{figure}
\begin{center}
\includegraphics{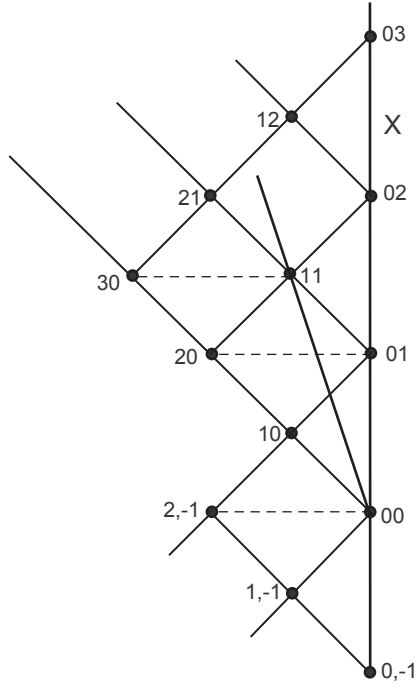}\caption{The causal construction of a timelike half cone. We have $x_{11}\in\Gamma_{1/3}^+(x_{00})$.}\label{quadrillage}
\end{center}
\end{figure}

Now choose a fraction $v=p/q$, with $0<p<q$. Let $n=p+q$ and $k=q-p$. Then $x_{n-k,k}\in\Gamma_{v}^+(x_{00})$. Moreover, if we let $x_{10}$ vary in $\Lambda_{00}^+$, then $x_{n-k,k}$ will vary in $\Gamma_{v}^+(x_{00})$. More precisely we have (writing $x=x_{00}$):

$$y\in\Gamma_{v}^+(x)\Leftrightarrow \exists x_{10}\in\Lambda_{x}^+, F(x_{10})$$

where $F(x_{10})$ is a first-order formula describing the construction of the lattice above, which uses only bound variables, half light-cones, colinearity, $X$, and $\Sigma_X$.

For instance, let us write the explicit formula in the case $v=1/3$:

\be 
y\in\Gamma_{1/3}(x)\Leftrightarrow\exists x_{10}\in\Lambda_x^+, (\exists x_{01}\in X\cap \Lambda_{x_{10}}^+, (\exists x_{20}\in(x_{10}x), x_{20}\Sigma_X x_{01}\wedge(y\in\Lambda_{x_{01}}^+\cap (x_{20}x_{02}))))
\ee

It is clear that we can easily generalize this construction to any positive rational number, and to $\Gamma_v^-$. We now extend it to irrational numbers in the following way. Let $v>0$ be  irrational. A point $y$ is in $\Gamma_v^+(x)$ if, and only if, for all $q\in{\mathbb Q}$ such that $q<v$, there exists $x'\in X$, with $x'<x$ and $y\in\Gamma_{q}^+(x')$, and for all $r\in {\mathbb Q}$ such that $r>v$, there exists $x''\in X$, with $x<x''$ and $y\in\Gamma_{r}^+(x'')$. We use a similar procedure for $\Gamma_v^-(x)$.

Finally, the relation $y\Gamma_{X,v}^\pm y'$ is equivalent to $\exists x\in X, y\in \Gamma_v^\pm(x)\wedge y'\in \Gamma_v^\pm(x)$. Thus, $\Gamma_{X,v}^\pm$ is $(X,\kappa,<)$-definable. This ends the proof.
\end{demo}

\section{Theorem 2}\label{secth2}
For convenience, we introduce an origin $O$ on $X$ and inertial coordinates on ${\mathcal M}$ adapted to $X$, that is, such that $X$ is the time axis.  It will be apparent that nothing in the constructions below depends on the choice of this particular system of coordinates, neither on the choice of $O$ on $X$. 

Let $f : [0;+\infty[\longrightarrow {\mathbb R}$ be any function such that $f(0)=0$. We denote by $\Gamma_{f}$ the set of all points of coordinates $(t,\vec{u})$ such that $t=f(\|\vec{u}\|)$. Thus $\Gamma_f$ is a hypersurface of revolution generated by the rotation around $X$ of the curve of the function $f$. It has just one point on $X$, which is the origin. If $\tau\in {\mathbb R}$, we denote by $f+\tau$ the translation by $\tau$ of the function $f$. Now define the relation $\Gamma_{X,\tilde{f}}$ by

\be
p\Gamma_{X,\tilde{f}}q\Leftrightarrow \exists\tau\in{\mathbb R}, p,q\in \Gamma_{X,f+\tau}
\ee

Recall that $\tilde{f}$ is the class of functions equal to $f$ up to an overall additive constant. Thus, we could equally have written $p\Gamma_{X,\tilde{f}}q\Leftrightarrow \exists f\in{\tilde f}, p,q\in \Gamma_{X,f}$.

\begin{figure}[hbtp]
\begin{center}
\includegraphics{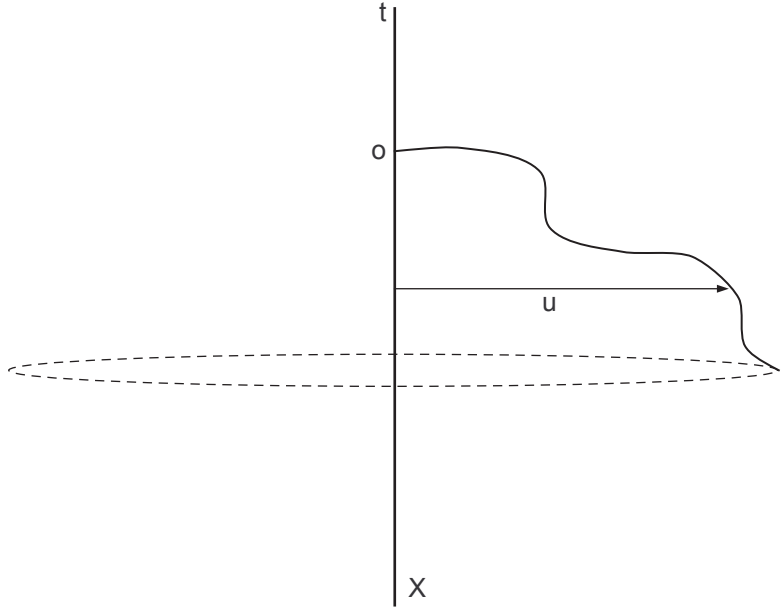}
\end{center}
\caption{The set $\Gamma_{X,f}$.}
\end{figure}

We see that the partition of ${\mathcal M}$ defined by the equivalence relation $\Gamma_{X,\tilde f}$ is given by the family of parallel hypersurfaces which are translated of $\Gamma_f$ in the direction of $X$. Note that $f$ need not be continuous.

\begin{theorem}\label{month2}
If $R$ is an equivalence relation on ${\mathcal M}$ which is $(X,\kappa,\tilde T)$-definable (or equivalently $(X,\eta,Or)$-definable), and which satisfies criterion (\ref{mycriterion}), then there exists $f : [0;+\infty[\longrightarrow {\mathbb R}$ such that $R=\Gamma_{X,\tilde f}$.
\end{theorem}
\begin{demo}
First, we can suppose that $T\in\tilde T$,  and the chosen inertial coordinate system are such that $T$ is the restriction of the time coordinate to $X$.  We will see in the end that nothing depends on the choice of $T$ in $\tilde T$. If $R$ is definable from $(X,\kappa,\tilde T)$, then it is invariant under automorphisms which preserves this structure, but this amounts to preserving $(X,\eta,Or)$. Thus $R$ is ${\rm Ilor}_X$-invariant.

Let $p=(0,\vec{u})$ be an event on $\Sigma_0$. By (\ref{mycriterion}) there exists a unique event $x_p$ on $X$ such that $pRx_p$. We write $t_p=T(x_p)$.

Since ${\rm Ilor}_X$ includes rotations of $\Sigma_0$ around $X$,  for any event $p'=(0,\vec{u'})\in\Sigma_0$ such that $\|\vec{u}\|=\|\vec{u'}\|$, we have $pRx_p\Rightarrow p'Rx_p$, that is $x_p=x_{p'}$. Thus, the mapping $f(r)=-T(x_p)$, where $x_p$ is any event in $\Sigma_0$ with space coordinates $\vec{u}$ such that $\|\vec{u}\|=r$, is a well defined function from $[0;+\infty[$ to ${\mathbb R}$ and it obviously satisfies $f(0)=0$.

Now take two events $p=(t_p,\vec{u})$ and $q=(t_q,\vec{v})$ such that $pRq$. By transitivity of $R$, one has $x_q=x_p$ (using notations like above). Let $\tau=T(x_q)=T(x_p)$. Define by $p'=(0,\vec{u})$ and $q'=(0,\vec{v})$. Since translations along $X$ belong to ${\rm Ilor}_X$, we have $pRx_p\Rightarrow p'Rx_{p'}$ with $T(x_{p'})=\tau-t_p$. Similarly, $T(x_{q'})=\tau-t_q$. Thus $f(\|\vec{u}\|)=-T(x_{p'})=t_p-\tau$ and $f(\vec{v})=-T(x_{q'})=t_q-\tau$. This implies that $p,q\in \Gamma_{X,f+\tau}$, that is, $p\Gamma_{X,\tilde f}q$.

Conversely, if $p\Gamma_{X,\tilde f}q$ then $p=(t_p,\vec{u})$ with $t_p=f(\|\vec{u}\|)+\tau$, and similarly, $q=(t_q,\vec{v})$ with $t_q=f(\|\vec{v}\|)+\tau$. Now, using translations along $X$, one has :

$$(0,\vec{u})R(-f(\|\vec{u}\|), \vec{0})$$
by definition, which entails

$$p=(t_p,\vec{u})R(t_p-f(\|\vec{u}\|), \vec{0})=(\tau,\vec{0})$$

Similarly, one finds $qR(\tau,\vec{0})$, thus $pRq$. We have proved that $R=\Gamma_{X,\tilde f}$. It is obvious that starting with $T+k$ instead of $T$, we would end up with a different $f$, but the same $\tilde f$.

We now need to show that each $\Gamma_{X,f}$ is $(X,\eta,T)$-definable. For this, take $p=(t_p,\vec{p})$. Define $\Psi^+(p)$ to be the intersection point of $\Lambda_p^+$ and $X$, and define similarly $\Psi^-(p)$ using $\Lambda^-_p$. Then $T(\Psi^+(p))=t_p+\|\vec{p}\|$ and $T(\Psi^-(p))=t_p-\|\vec{p}\|$.

To prove that $\Gamma_{X,f}$ is $(X,\preceq,T)$-definable, it is sufficient, using transitivity, to prove that the formula $p\Gamma_{X,\tilde f}x$ is so definable.

Now

\be
p\Gamma_{X,\tilde f}x\Leftrightarrow  T(x)=t_p-f(\|\vec{p}\|)={T(\Psi^+(p))+T(\Psi^-(p))\over 2}-f({T(\Psi^+(p))-T(\Psi^-(p))\over 2})
\ee\label{defloc}

This proves the result.
\end{demo}
\section{Conclusion}

With theorem \ref{month2}, we have come close to constructing an arbitrary time coordinate. There are some differences though: there is no smoothness requirement on the simultaneity classes, they are not either required to be spacelike, and finally they are necessarily spatially symmetric. This last feature can be easily traced back to the tools we are given to build our relation: a clock, the spherical wavefronts of light emitted from $X$, and the (god-given) causal connectibility relation $\kappa$. We could quite simply break this symmetry by using directional laser pointers instead of spherically symmetric light emmiters.

However, we can wonder if it legitimate to use $\kappa$ in the initial data, given that $\kappa$ cannot be recovered from the information available to $X$. It would certainly be more adequate to use 

\begin{itemize}
\item $X$ itself, as usual,
\item The set $L_X$ of all light cones $\Lambda_x$ with $x\in X$,
\item perhaps some additional structures on $X$ such as $<$ or $T$.
\end{itemize}

The following result  is a corollary of theorem \ref{montheorem1}:

\begin{theorem}\label{month3}
\begin{itemize}
\item There is no implicitly $(X,L_X)$-definable equivalence relation on ${\mathcal M}$ satisfying (\ref{malamcond}).
\item The only implicitly $(X,L_X,<)$-definable equivalence relations satisfying (\ref{malamcond}) are $\Lambda_X^+$ and $\Lambda_X^-$. Conversely, these two relations are explicitly $(X,L_X,<)$-definable.
\item The relations $\Gamma_{X,f}$ are all explicitly $(X,L_X,\tilde T)$-definable.
\end{itemize}
\end{theorem}

First, implicitly $(X,L_X)$-definable relations are ${\rm Aut}_{X,\kappa}$-invariant, and thus, by Malament's theorem, there can only be $\Sigma_X$ among the non-trivial ones. It remains to show that $\Sigma_X$ is not ${\rm Aut}_{X,L_X}$-invariant. The idea is to use a non-affine causal automorphism of a Lorentzian $2$-plane (this sort of thing exists in dimension $2$ but not in higher dimensions, as shown in \cite{zeeman}), and propagate it by rotational symmetry. This will leave $L_X$ invariant (and tear apart the other light-cones). More precesily, let $f$ be a bijective map from $X$ to $X$, and let $\phi_f : {\mathcal M}\rightarrow {\mathcal M}$ be defined, in inertial coordinates, by :

\be
\phi_f(t,\vec{u})=({1\over 2}(f(t+\|\vec{u}\|)+f(t-\|\vec{u}\|)),{1\over 2}|f(t+\|\vec{u}\|)-f(t-\|\vec{u}\|)|{\vec{u}\over\|\vec{u}\|})
\ee

for $\vec{u}\not=\vec{0}$, that is $(t,\vec{u})\notin X$, and $\phi(t,\vec{0})=f(t)$. We let the reader check that $\phi_f$ is a bijection such that :

\begin{itemize}
\item $\phi_f(X)=X$,
\item $\forall x\in X$, $\phi_f(\Lambda_x)=\Lambda_{\phi_f(x)}$.
\end{itemize}

Thus it is an element\footnote{It is an interesting exercise to prove that $\psi\in{\rm Aut}_{X,L_X}$ if and only if $\psi=\phi_f\circ T\circ \tau$, where $\phi_f$ is as above, $T$ is a ``twist'', that is a bijection of ${\mathcal M}$ to itself which globally preserves all the $3$-spheres centered on $X$, and $\tau$ is a translation in the direction of $X$.} of ${\rm Aut}_{X,L_X}$. However, if we take a non-affine bijection $f$, for instance $f(t)=t^3$, it does not preserve $\Sigma_X$.

Now choosing an $f$ which is increasing with respect to $<$, and non-linear (for instance $f(t)=t^3$ in adapted coordinates), we see immediately that the only $\phi_f$-invariant relations listed in theorem \ref{montheorem1} are $\Lambda_X^+$ and $\Lambda_X^-$. The converse part is obvious.

Finally, to show the last statements we only need to remark that in order to prove the converse part of theorem \ref{month2}, we needed $\kappa$ at two points: when we use $\Sigma_X$ to define $t_p$ and when we define $\Psi^\pm(p)$. But, using the procedure described at the beginning of this paper, it is easy to define $\Sigma_X$ using $L_X$ and $\tilde T$. Moreover, $\Psi^+(p)$ can be defined as the unique point in $X$ such that $p\in\Lambda_{\Psi^(p)}^-$, and similarly for $\Psi^-(p)$. In this way, formula (\ref{defloc}) uses only $L_X$ and $\tilde T$.

It is interesting to note that when one restricts from the global $\kappa$ to the ``local'' structure $L_X$, then $\Sigma_X$, which was unique in the first setting, disappears in the second. Of course it reappears as soon as one introduces $T$, but it is then no longer unique. The only point where some sort of unicity now arises is in the second statement of theorem \ref{month3}. Maybe this could be used to argue that $\Lambda_X^\pm$ are ``less conventional'' than $\Sigma_X$ from a local point of view.

\bibliographystyle{alpha}
\bibliography{bibliosimul}

\end{document}